\documentclass{article}

\usepackage{PRIMEarxiv}

\usepackage[utf8]{inputenc} 
\usepackage[T1]{fontenc}    
\usepackage{hyperref}       
\usepackage{url}            
\usepackage{booktabs}       
\usepackage{amsfonts}       
\usepackage{nicefrac}       
\usepackage{microtype}      
\usepackage{lipsum}
\usepackage{fancyhdr}       
\usepackage{graphicx}       
\graphicspath{{media/}}     

\usepackage{setspace}   
\usepackage[numbers]{natbib}

\usepackage{multirow} 
\usepackage{stfloats} 
\usepackage{xcolor} 

\definecolor{darkishred}{rgb}{1,0.3,0.3}
\definecolor{darkgreen}{rgb}{0,0.3,0}
\definecolor{checkmarkgreen}{rgb}{0,0.8,0}
\definecolor{colorFluidity1}{rgb}{1,0.6,0}
\definecolor{colorFluidity2}{rgb}{0.70,0.85,0}
\definecolor{colorFluidity3}{rgb}{0.0,0.8,0}
\newcommand{\stableSystem}[0]{\textcolor{checkmarkgreen}{\checkmark}}
\newcommand{\crashedSystem}[0]{\textcolor{red}{X}}
\newcommand{\fluidityZero}[0]{\textcolor{red}{-}}
\newcommand{\fluidityOne}[0]{\textcolor{colorFluidity1}{+}}
\newcommand{\fluidityTwo}[0]{\textcolor{colorFluidity2}{++}}
\newcommand{\fluidityThree}[0]{\textcolor{colorFluidity3}{+++}}

\title{VTX: Real-time high-performance molecular structure and dynamics visualization software
\thanks{\textit{\underline{Citation}}: 
\textbf{Maxime Maria, Simon Guionnière, Nicolas Dacquay, Cyprien Plateau–Holleville, Valentin Guillaume, Vincent Larroque, Jean Lardé, Yassine Naimi, Jean-Philip Piquemal, Guillaume Levieux, Nathalie Lagarde, Stéphane Mérillou and Matthieu Montes. VTX: Real-time high-performance molecular structure and dynamics visualization software. Bioinformatics, btaf295 DOI:10.1093/bioinformatics/btaf295.}} 
}

\begin{document}
\maketitle

Maxime Maria \textsuperscript{1}, Simon Guionnière \textsuperscript{2}, Nicolas Dacquay \textsuperscript{2}, Cyprien Plateau–Holleville \textsuperscript{1}, Valentin Guillaume \textsuperscript{2}, Vincent Larroque \textsuperscript{1,3}, Jean Lardé \textsuperscript{2,3}, Yassine Naimi \textsuperscript{3}, Jean-Philip Piquemal  \textsuperscript{4,5,6}, Guillaume Levieux \textsuperscript{7}, Nathalie Lagarde \textsuperscript{2}, Stéphane Mérillou \textsuperscript{1} and Matthieu Montes \textsuperscript{2,6,*}

\begin{spacing}{0.75}
\begingroup
    \tiny 
    \textsuperscript{1}XLIM, UMR CNRS 7252, Université de Limoges, 87000, Limoges, France, 
    \textsuperscript{2}Laboratoire GBCM, EA 7528, Conservatoire National des Artset Métiers, 75003, Paris, France, 
    \textsuperscript{3}Qubit Pharmaceuticals SAS, France, 
    \textsuperscript{4}LCT, UMR 7616 CNRS, Sorbonne Université, Paris, France,
    \textsuperscript{5}Department of Biomedical Engineering, University of Texas at Austin, Texas, USA, 
    \textsuperscript{6}Institut Universitaire de France, Paris, France and
    \textsuperscript{7}Laboratoire CEDRIC, EA 4626, Conservatoire National des Arts et Métiers, 75003, Paris, France.\\
    *Corresponding authors: maxime.maria@unilim.fr, matthieu.montes@cnam.fr\\
\endgroup
\end{spacing}

\begin{abstract}
    \textbf{Summary:} VTX is a molecular visualization software capable to handle most molecular structures and dynamics trajectories file formats. It features a real-time high-performance molecular graphics engine, based on modern OpenGL, optimized for the visualization of massive molecular systems and molecular dynamics trajectories. VTX includes multiple interactive camera and user interaction features, notably free-fly navigation and a fully modular graphical user interface designed for increased usability. It allows the production of high-resolution images for presentations and posters with custom background. VTX design is focused on performance and usability for research, teaching and educative purposes.\\
\textbf{Availability and implementation:} VTX is open source and free for non commercial use. Builds for Windows and Ubuntu Linux are available at http://vtx.drugdesign.fr. The source code is available at https://github.com/VTX-Molecular-Visualization.\\
\textbf{Contact:}maxime.maria@unilim.fr; matthieu.montes@cnam.fr\\
\textbf{Supplementary Information:} A video displaying free-fly navigation in a whole-cell model is available
\end{abstract}


\section{Introduction}

Molecular visualization is a critical task usually performed by structural biologists and bioinformaticians to aid the different processes that are essential to understand structural molecular biology \citep{Olson18}. 
Following the recent advances in the determination of atomic resolution molecular structures and assemblies \citep{cryoem}, in protein structure prediction \citep{alphafold,alphafolddb}, and in the increased accessibility of molecular dynamics simulation \citep{amaro24,hospital24}, there is a profusion of molecular structural biology data that is available to the scientific community. 
Due to the modern High-Performance Computing hardware and storage, the size of the simulated systems~\citep{sanbonmatsu,marrink23,aksimentiev24} with molecular dynamics has dramatically increased and the storage, analysis and interactive visualization of the resulting data can become problematic for currently available molecular visualization systems such as Py\MakeUppercase{mol} \citep{pymol}, VMD \citep{vmd} or Chimera-X \citep{chimeraX}. 

Here, we present VTX, an open-source molecular visualization software. VTX is optimized to handle efficiently and in real-time the big data in molecular simulations notably by including a meshless high performance molecular graphics engine coupled with a minimalistic task-oriented GUI to maximize the usability for non-expert users. 
VTX uses the chemfiles library \citep{chemfiles} to read and write molecular data which handles most widely used molecular structures and trajectories file formats. VTX includes various representations and rendering options and provides different tools such as structural alignment and distance/angle measurement for interactive analysis of massive molecular scenes. It is free and open source for non commercial use and available on linux and windows at \href{http://vtx.drugdesign.fr}{http://vtx.drugdesign.fr} and \href{https://github.com/VTX-Molecular-Visualization}{https://github.com/VTX-Molecular-Visualization}

\section{Methods}\label{sec:method}

\subsection*{Molecular Graphics Engine}
Thanks to adapted data structures and rendering algorithms, VTX is designed for the real-time visualization of very large molecular systems, composed of several millions atoms, on a consumer laptop. %
A molecule is defined by the position of its atoms, their corresponding radius and connectivity.

\textit{\textbf{Meshless Representations:}}

Most of VTX representations are described implicitly, without using a triangular mesh (sticks, ball and sticks, Van der Waals, Solvent Accessible Surface).
These meshless representations allow impostor-based techniques in the rendering engine \citep{impostors}.  For each primitive, a simple quad is rasterized. Then, ray-casting is used to evaluate the implicit equation of the primitive and display the final shape.
This allows fast and pixel perfect rendering while reducing the memory consumption and bandwidth usage which is essential to handle large molecular structure or dynamics data.

\textit{\textbf{Cartoon and SES representations:}}
The Cartoon representation is generated on-the-fly and displayed with adaptive level-of-detail (LOD) method using tessellation shaders \citep{cartoonSS}.%
The Solvent Excluded Surface is computed following a discrete approach, similarly to \citep{lindow14} and the resulting surface is extracted via marching cubes \citep{marchingcubes}.%

\begin{figure*}[!t]
    \centering 
    
    \includegraphics[width=0.99\textwidth]{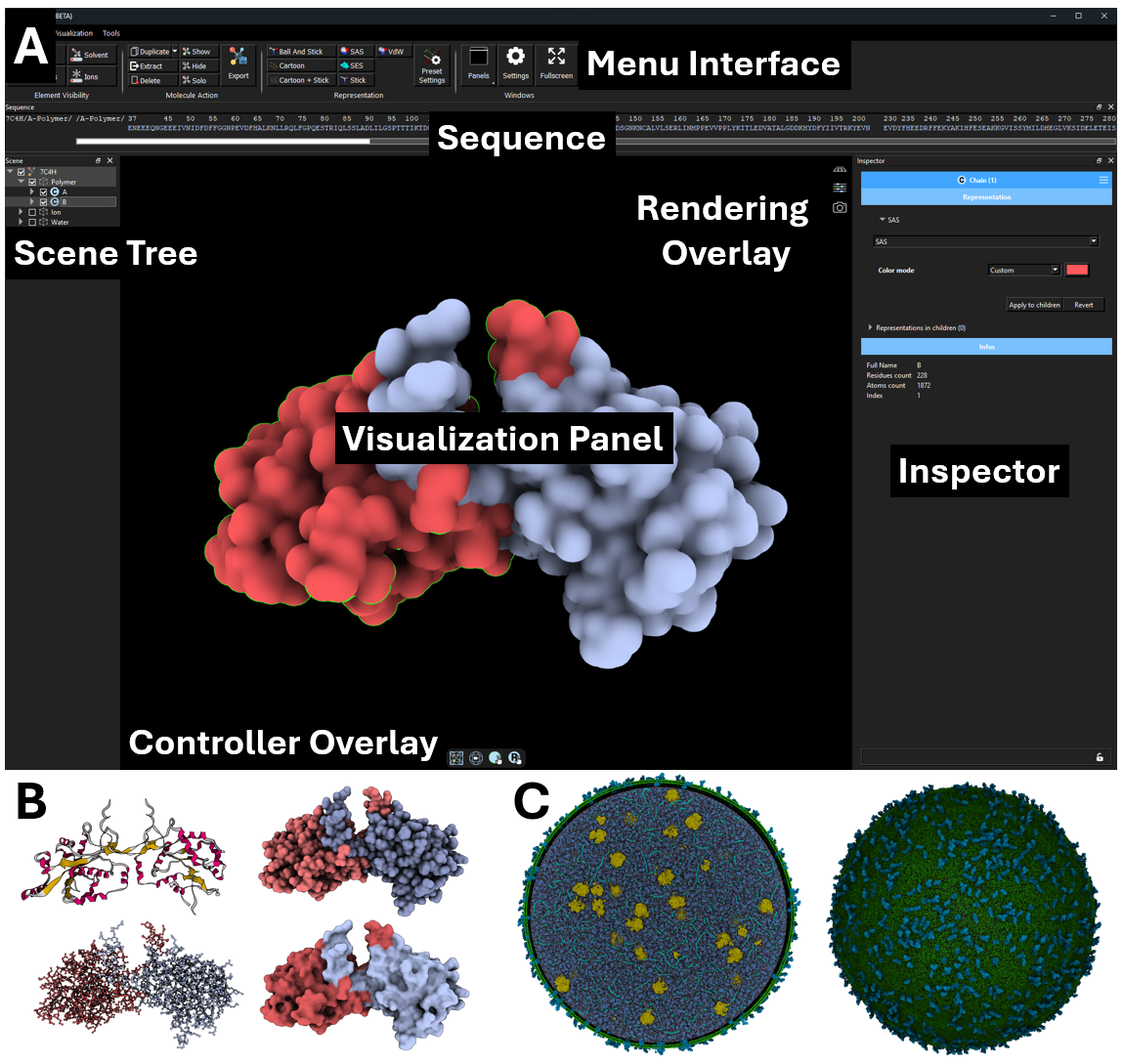}
   
    \caption{\textbf{A.} Illustration of the VTX GUI organized with different dockable panels. The \textit{Scene Tree} panel contains and allows the selection of all the objects present in the scene, such as molecules, labels and viewpoints. The \textit{Sequence} panel displays and allows the selection of the sequence of each biomolecule , while the \textit{Visualization Panel} allows the selection, observation, and manipulation of their 3D structure. Frequently used options for manipulation and rendering are directly accessible in the visualization panel through the use of button overlays. When an object is selected in the scene, detailed information about the loaded object, additionally loaded trajectories and chosen representations are available in the \textit{Inspector} panel. The \textit{Console} panel displays all logs. \textbf{B.} Different molecular representations available in VTX (Saccharomyces Cerevisiae BCP1, PDB id~: 7C4H) Cartoon (upper left), Van der Waals (upper right), balls and sticks (bottom left), Solvent Excluded Surface (bottom right) \textbf{C.} Illustration of the coarse-grained whole mycoplasma cell model from \citep{marrink23}  with (left) and without (right) clipping. Martini beads are displayed as surface and flat color (membrane lipids in green, membrane proteins in blue, ribosomes in yellow, chromosome in light blue, cytoplasmic proteins in dark blue).}
    \label{fig:figure1}
\end{figure*}

\textit{\textbf{High Quality Rendering:}} 
VTX aims to enhance the user’s visual analysis experience in real-time by providing high-quality rendering Figure (~\ref{fig:figure1}B). The use of meshless representations in VTX results in a pixel-perfect display quality.
Additionally, VTX offers various rendering options to improve the perception of details and enable the creation of visually appealing illustrations. The deferred rendering approach allows for the implementation of various post-processing techniques, including shading (flat, matte, glossy, toon), ambient occlusion, fog, outline, and anti-aliasing. These techniques can enhance the visual perception of molecular shape and improve overall image quality. VTX offers an image export feature that allows to produce an image of the displayed molecular scene with a resolution up to 8K and a custom background with user-defined transparency. %
When dealing with dynamic data (such as MD trajectories), ambient occlusion rendering cannot be pre-computed. In VTX, we use a deferred shading graphics pipeline to compute lighting on surfaces and apply screen-space post-processes that allows to produce high-quality rendering in real time \citep{mittring07}.

\subsection*{Improved user experience}

\textit{\textbf{Camera modes:}} VTX disposes of an interactive camera system controllable via the keyboard and/or mouse that includes different modes: 1. trackball and 2. free-fly. The trackball mode is the classical camera  available in molecular visualization software where the camera revolves around a fixed focus point. This allows the user to rotate the view around the object of interest. The first-person free-fly navigation mode allows the user to fully control the movement of the camera. This mode is similar to the first-person perspective in video games where the user can move freely in the 3D space. It has notably been used in UDock \citep{levieux2014, udock2} and in the multiscale molecular rendering tool CellVIEW \citep{cellview} 
\\
\textit{\textbf{Keyboard controls:}} Using a trackpad to navigate in 3D molecular scenes can be challenging because it may not provide the same level of precision and control as a mouse. This can make it difficult to perform precise tasks such as rotating, zooming, and panning the view. VTX provides alternative navigation controls using keyboard bindings to make it easier to use with a trackpad and to complement the mouse for the freefly navigation mode. 
\\
\textit{\textbf{Customizable Graphical User Interface (GUI):}} The GUI of VTX enables to create customizable presets of combined displays and representations that can be saved and easily accessed. The different dockable windows of the software can be moved, resized, and arranged according to the user’s preferences, allowing for a customizable workspace. It also includes quick access to the most frequently used commands in the 3D visualization window through the use of clearly labeled buttons.

\subsection*{File formats compatibility}
The Chemfiles library \citep{chemfiles} is used to handle diverse file formats for molecular structures (including PDB, mol2, mmCIF and MMTF) 
and molecular dynamics trajectories (including dcd, xtc and arc).
While mmCIF is the preferred format in VTX for molecular data as its structure is well adapted for very large molecular structures and assemblies, a direct download access to the PDB \citep{Berman2000} is provided through the use of the PDB API.

\subsection*{Implementation}
VTX is written in C++, with modern OpenGL for hardware-accelerated rendering and the Qt \citep{qt} framework for its graphical user interface (GUI).

\subsection*{Operation}
VTX operates on computers supporting OpenGL 4.5. The software is compatible with Windows 10 and subsequent versions, as well as Ubuntu Linux versions 20 and higher.

\section*{Performance evaluation}

The performance of VTX version 0.4.4 in terms of system file reading and fluidity of manipulation has been comparatively evaluated to VMD version 1.9.4 \citep{vmd}, Py\MakeUppercase{mol} version 3.1 \citep{pymol} and ChimeraX version 1.9 \citep{chimeraX}.  The performance of each software was assessed using a benchmarking dataset comprising diverse molecular systems, characterized by varying sizes ranging from 100,000 atoms to 100 million Martini beads. The evaluation focused on two primary metrics: 1. system stability, defined as the software's ability to load the molecular system without crashing; and 2. interactive fluidity, quantified by the smoothness and responsiveness of user interaction during selection/change of representation and manipulation tasks, graded using different levels (\fluidityZero crashed or frozen, \fluidityOne low fluidity, \fluidityTwo moderate fluidity, and \fluidityThree high fluidity).

The benchmarking dataset is comprised of four different systems: 1. the structure of type IVa pilus machine \citep{3jc8} containing 107,640 atoms (PDB ID : 3JC8); 2. the structure of the P68 bacteriophage \citep{6q3g} containing 1,074,183 atoms (PDB ID : 6Q3G); 3. a concatenated model involving 3 structures: 2 microtubules \citep{8J078GLV} (PDB ID : 8J07 and 8GLV) and a reconstruction of a bacteroides phage crAss001 \citep{8ckb} (PDB ID : 8CKB) for a total of 11,003,162 atoms; and 4. the 2023 Martini minimal whole cell model from \citep{marrink23} displayed in figure \ref{fig:figure1}C that contains 60,887 soluble proteins, 2,200 membrane proteins, 503 ribosomes, a single 500 kbp circular dsDNA, 1.3 million lipids, 1.7 million metabolites and 14 million ions for a total of 101,753,154 Martini beads. All evaluations have been performed on a Dell Alienware 15r with an intel i7-10750H CPU and a RTX2080 Super. Resulting performance is displayed in table \ref{table:bench}. %

No values could be obtained with ChimeraX and Pymol on the 100M system, due respectively to a crash and a freeze during system loading. Similarily, even though VMD could open 10M and 100M systems, the selection and manipulation of the system resulted in a freeze. Despite limited computational resource, VTX allowed to easily manipulate the larger systems in real-time, perform precise selections and modify rendering settings. The free-fly real-time navigation video presented in supplementary material was captured with microsoft XBOX overlay on windows 11. 

\begin{table*}[]
    \centering
    \resizebox{\columnwidth}{!}{%
    \begin{tabular}{c|cccc|cccc|}
    \cline{2-9}
                           & \multicolumn{4}{c|}{System Stability}                                                    & \multicolumn{4}{c|}{Interactive Fluidity}                                                    \\ \cline{2-9} 
                           & \multicolumn{1}{c|}{ ChimeraX} & \multicolumn{1}{c|}{PyMOL} & \multicolumn{1}{c|}{VMD} & VTX & \multicolumn{1}{c|}{ ChimeraX} & \multicolumn{1}{c|}{PyMOL} & \multicolumn{1}{c|}{VMD} & VTX \\ \hline
    \multicolumn{1}{|c|}{3jc8 (107,640 atoms)} & \multicolumn{1}{c|}{\stableSystem} & \multicolumn{1}{c|}{\stableSystem} & \multicolumn{1}{c|}{\stableSystem} & \stableSystem & \multicolumn{1}{c|}{\fluidityThree} & \multicolumn{1}{c|}{\fluidityThree} & \multicolumn{1}{c|}{\fluidityThree} & \fluidityThree \\ \hline
    \multicolumn{1}{|c|}{6q3g (1,074,183 atoms)} & \multicolumn{1}{c|}{\stableSystem} & \multicolumn{1}{c|}{\stableSystem} & \multicolumn{1}{c|}{\stableSystem} & \stableSystem & \multicolumn{1}{c|}{\fluidityThree} & \multicolumn{1}{c|}{\fluidityThree} & \multicolumn{1}{c|}{\fluidityTwo} & \fluidityThree \\ \hline
    \multicolumn{1}{|c|}{8ckb 8j07 8glv (11,003,162 atoms) } & \multicolumn{1}{c|}{\stableSystem} & \multicolumn{1}{c|}{\stableSystem} & \multicolumn{1}{c|}{\stableSystem} & \stableSystem & \multicolumn{1}{c|}{\fluidityOne} & \multicolumn{1}{c|}{\fluidityThree} & \multicolumn{1}{c|}{\fluidityZero} & \fluidityThree \\ \hline
    \multicolumn{1}{|l|}{Whole cell model (101,753,154 Martini beads)} & \multicolumn{1}{c|}{\crashedSystem} & \multicolumn{1}{c|}{\crashedSystem} & \multicolumn{1}{c|}{\stableSystem} & \stableSystem & \multicolumn{1}{c|}{\fluidityZero} & \multicolumn{1}{c|}{\fluidityZero} & \multicolumn{1}{c|}{\fluidityZero} & \fluidityTwo \\ \hline
    \end{tabular}%
    }   
        \caption{Comparative performance evaluation of reference molecular visualization software with VTX on different molecular systems with sizes ranging from 100K atoms to 100 million Martini beads. System stability measures the ability of the software to load the given system without a crash (\stableSystem successful loading, X crashed). Interactive stability measures the smoothness and responsiveness of user interaction during selection, change of representation and manipulation tasks, graded using different levels (- crashed or frozen, + low fluidity, ++ moderate fluidity, and +++ high fluidity). All evaluations have been performed on a Dell Alienware 15r with an intel i7-10750H CPU and a RTX2080 Super.
        }
        \begingroup
        \endgroup
        \label{table:bench}
\end{table*}

\section*{Use case}
The VTX GUI is organized as presented in Figure~\ref{fig:figure1}A. Molecular structure and trajectories can be loaded using the \textit{open} button from the file menu or directly downloaded from the PDB with their accession number. For ease of use, molecular trajectories can be loaded with a right click onto an already loaded molecular object in the scene tree.  

As illustrated in Figure~\ref{fig:figure1}C, due to its new generation molecular graphics engine, VTX is highly scalable and allows the rendering in real time of massive molecular systems (100+ million Martini beads) on a consumer laptop with a NVIDIA RTX2080m. 

VTX also includes different functions in the \textit{tools} menu tab for distance and angle measurements, as well as structural alignment using the CE method \citep{Shindyalov1998}. 
The session state, including the user's organization of the VTX workspace, can be saved and exported. High resolution illustrations with or without background can be generated with the \textit{snapshot} function.

\section*{Conclusion and perspectives}

VTX is a molecular visualization software designed to provide:
   1. a high-quality real time rendering of large molecular systems (several hundred million atoms on a consumer laptop) 2. a comfortable user experience with intuitive controls and tools and 3. a wide compatibility with most molecular structures and trajectories file formats and a high-quality image export feature allowing to produce up to 8K-resolution images with custom background for posters and presentations. Future versions will include real-time analytical SES \citep{SEScyprien}, high-end real-time ambient occlusion, offline ray-tracer for high-quality illustrations and movies rendering, python-like command binding. VTX is open source and free for non commercial use.

\section*{Data and software availability} 

\subsection*{Data availability}
\sloppy 
All files composing the benchmarking dataset, except the Martini minimal whole cell model, are available at \href{https://doi.org/10.5281/zenodo.14962673}{https://doi.org/10.5281/zenodo.14962673}

All .gro files that constitute the Martini minimal whole cell model are available upon request at the Marrink lab. The procedure to generate the model is available at  \href{https://github.com/marrink-lab/Martini\_Minimal\_Cell}{https://github.com/marrink-lab/Martini\_Minimal\_Cell}

\subsection*{Software availability}

VTX is open source and free for non commercial use under the VTX consortium license. Builds for Windows and Linux are available at http://vtx.drugdesign.fr. The source code is available at https://github.com/VTX-Molecular-Visualization.

\section{Competing interests}
No competing interest is declared.

\section{Author contributions statement}
MMa, SG and MMo designed the software; MMa developed the first version of the rendering engine. MMa, SG, ND, CPH, VG, VL, JL, YN contributed to the code; ND, JL and GL designed the interface; MMa, VG and MMo wrote the manuscript; MMa, VG, JPP, NL, SM, GL and MMo reviewed the manuscript

\section{Acknowledgments}
We thank Prof. SJ Marrink and J Stevens for kindly providing test data of massive molecular systems. We thank all beta testers for their time and fruitful feedback. 
VTX has received funding from the European Research Council (ERC) under the European Union's Horizon 2020 research and innovation program (grant agreement n° 640283) and from the ERC under the Horizon Europe research and innovation program (grant agreement n° 101069190). CPH and VL are supported by institutional grants
from the National Research Agency under the Investments for the
future program with the reference ANR-18-EURE-0017 TACTIC.
VL is receipent of a fellowship from Qubit Pharmaceuticals. 

\pagebreak
\bibliographystyle{unsrt}  
\bibliography{article}

\end{document}